\documentclass[11pt]{article}
\usepackage{a4wide}                      % to get a wider page margen
\usepackage{epsf}                        % to make figures
\usepackage{latexsym}                    % to make fancy symbols
\usepackage{amsfonts}                    % to make fancy symbols
\usepackage{amssymb}                     % to make fancy symbols
\usepackage{latexsym}                    % to make fancy symbols
\usepackage{amsfonts}                    % to make fancy symbols
\usepackage{amssymb}                     % to make fancy symbols
\def\Z{\ensuremath{\mathbb{Z}}}
\def\R{\ensuremath{\mathbb{R}}}

\def\makeatletter{\catcode`\@=11}% 11:letter
\makeatletter
\def\mathbox#1{\hbox{$\m@th#1$}}%
\def\math@ccstyles#1#2#3#4#5#6#7{{\leavevmode
      \setbox0\mathbox{#6#7}%
      \setbox2\mathbox{#4#5}%
      \dimen@ #3%
      \baselineskip\z@\lineskiplimit#1\lineskip\z@
      \vbox{\ialign{##\crcr
             \hfil \kern #2\box2 \hfil\crcr
             \noalign{\kern\dimen@}%
             \hfil\box0\hfil\crcr}}}}
\def\mathaccstyles{\math@ccstyles\maxdimen}
\def\maththroughstyles{\math@ccstyles{-\maxdimen}}
\def\unity%
{\maththroughstyles{.45\ht0}\z@\displaystyle {\mathchar"006C}\displaystyle 1}

\begin{document}

\rightline{FFUOV-07/01}
\rightline{hep-th/0701133}
\rightline{January 2007}
%\vspace{2.5cm}
\vspace{3.5cm}

%%%%%%%%%%%%%%%%%
\hspace*{-0.3cm}
\centerline{\LARGE \bf Pp-wave Matrix Models from Point-like Gravitons}\footnote{Talk presented by Y.L. at the 2nd TMR workshop, Napoli, Oct. 9-13, 2007}
%\vspace{1.3truecm}
\vspace{1.5truecm}

\centerline{
    {\large \bf Yolanda Lozano${}^{a,}$}\footnote{E-mail address:
                                  {\tt yolanda@string1.ciencias.uniovi.es}}
    {\bf and}
    {\large \bf Diego Rodr\'{\i}guez-G\'omez${}^{b,}$}\footnote{E-mail address:
                                  {\tt drodrigu@Princeton.EDU}}
                                                   }
\vspace{.4cm}
\centerline{{\it ${}^a$Departamento de F{\'\i}sica,  Universidad de Oviedo,}}
\centerline{{\it Avda.~Calvo Sotelo 18, 33007 Oviedo, Spain}}

\begin{center}
\centerline{{\it ${}^b$Department of Physics, Princeton University, }}
\centerline{{\it Princeton, NJ 08540, USA}}
\end{center}

%\vspace{2truecm}
\vspace{2.5truecm}
%%%%%%%%%%%%%%%%%
\centerline{\bf ABSTRACT}
\vspace{.5truecm}

\noindent
The BFSS Matrix model can be
regarded as a theory of 
coincident M-theory gravitons. In this spirit, we summarize how  using the action for coincident gravitons
proposed in hep-th/0207199
it is possible to go beyond the linear order approximation of Kabat and Taylor, and to provide a satisfactory microscopical description of giant gravitons  
in $AdS_m\times S^n$ backgrounds.
We then show that in
the M-theory maximally supersymmetric pp-wave background, the action for coincident gravitons, besides 
reproducing the BMN Matrix model, predicts
a new quadrupolar coupling to the M-theory 6-form potential, which supports the so far elusive
fuzzy 5-sphere giant graviton solution. Finally, we discuss 
similar Matrix models that can be derived in Type II string theories using dualities.

\newpage

\section{Introduction}

Several approaches have been taken in the literature for the study of M-theory and Type II pp-wave
Matrix models \cite{BMN,DSV,SY,DMS,SJ}.
The derivation of the BMN Matrix model was based on the generalization of the action for a superparticle in the pp-wave background to an arbitrary number of particles, using the
requirement of
consistency with supersymmetry \cite{BMN}. 
The same Matrix model was obtained in \cite{DSV}
by regularizing the light-cone supermembrane action in the pp-wave
background. In turn, Matrix String theory in Type IIA pp-wave backgrounds has been studied in \cite{SY,DMS}\footnote{See also \cite{Bonelli}.}, by either starting from 
the supermembrane
action in the maximally supersymmetric pp-wave background of M-theory and using 
the correspondence law of \cite{SY2} to reduce it to ten dimensions, or by constructing it
from the BMN Matrix action
using the 9-11 flip. In Type IIB, general features about a Matrix String theory in the maximally supersymmetric pp-wave background
have been discussed in \cite{Gopakumar}, and an explicit Matrix String theory for this
background has been constructed in \cite{SJ}.
The  approach taken in \cite{SJ} was to regularize the light-cone 3-brane action in the 
Type IIB pp-wave background, in the same spirit of
\cite{DSV}.
The light-cone 3-brane carries $N$ units of light-cone momentum, and some of its vacua are finite
size 3-branes with zero light-cone energy, i.e. giant
gravitons \cite{SS}.
In close analogy to the giant graviton description in \cite{BMN}, reference \cite{SJ} proposes
a description of the 3-sphere  vacua in terms of $N$ expanding
gravitons, each carrying one unit of light-cone momentum. These are the so-called
{\it tiny} gravitons. 

We show in this note that both the BMN Matrix model and the Tiny Graviton Matrix Theory of \cite{SJ} can be regarded as theories of coincident gravitons, expanding by Myers dielectric effect into their corresponding giant graviton vacua. {}From this point of view the tiny gravitons of reference \cite{SJ} are simply
Type IIB point-like gravitons. 

We start in section 2 by reviewing some properties of the action for coincident M-theory gravitons constructed in \cite{JL2}. We also recall some of our results regarding
the microscopical description of giant graviton configurations in $AdS_m\times S^n$ spacetimes using this action. Section 3 is devoted to the derivation of pp-wave Matrix models from the action for point-like gravitons. We start in subsection 3.1 with the derivation of the BMN Matrix model. We show that the elusive 5-sphere giant graviton solution can be reproduced thanks to a new coupling to the 6-form potential present in our Matrix model. In subsection 3.2 we derive a new Matrix model in the Type IIB pp-wave background which supports fuzzy 3-sphere giant graviton solutions with the right behaviour in the large $N$ limit. We briefly discuss some relations between this Matrix model and the Tiny Graviton Matrix theory of \cite{SJ}. Due to lack of space we refer the reader to reference \cite{LR2} for the explicit derivation of the Matrix String theory in the Type IIA background of \cite{SY,DMS} from the action for coincident Type IIA gravitons.

\section{The action for M-theory gravitons}

The worldvolume theory associated to $N$ coincident gravitons in M-theory
 is a $U(N)$
 gauge theory, in which the vector
 field is associated to M2-branes (wrapped on the direction of propagation of the waves)
 ending on them \cite{JL2}. This vector field gives the BI field living in a set of coincident 
 D0-branes upon reduction along the direction of propagation of the
 waves. 
 
 In this note we will use a truncated version of the action in \cite{JL2}
 in which the vector field is set to zero. This action is
given by 
\begin{eqnarray}
S&=&- \int d\tau \ \mbox{STr} 
    \left\{ k^{-1}\sqrt{-P[E_{00}+E_{0i}(Q^{-1}-\delta)^i_k E^{kj}E_{j0}]\ 
                    \det Q} \ \right\}
\nonumber
\\ 
&+& \int d\tau \ \mbox{STr} \left\{ -P[k^{-2} k^{(1)}]
                    +i P[(\mbox{i}_X \mbox{i}_X)C^{(3)}] +
                    \textstyle{\frac12} P[(\mbox{i}_X \mbox{i}_X)^2 \mbox{i}_k C^{(6)}] 
                    +\cdots  \right\}, 
\label{11daction}\\
&& \hspace*{-.8cm} 
E_{\mu\nu}= g_{\mu\nu}- k^{-2}k_\mu k_\nu +k^{-1}(\mbox{i}_k C^{(3)})_{\mu\nu}, \nonumber
\hspace{2cm}
Q^i_j=\delta^i_j + ik[X^i,X^k]E_{kj}\, ,
\end{eqnarray}
where we have set the tension (the momentum charge) of a single graviton to one. 
This action contains the
direction of propagation of the waves as a special isometric direction, with
Killing vector $k^\mu$. In the Abelian limit, a Legendre transformation restoring the
dependence on this direction yields the usual action for massless particles. In turn,
in the non-Abelian case, this action gives rise to Myers action for coincident D0-branes
after dimensional reduction over $k^\mu$. Notice that the waves are minimally coupled to the momentum operator
$k^{(1)}_\mu/k^2=g_{z\mu}/g_{zz}$, in coordinates adapted to the isometry in
which $k^\mu=\delta^\mu_z$. The reader is referred to references \cite{JL2,JL1} for more details about the construction of this action.

An important check of the validity of this action is that it has been successfully used in the
 microscopical study of giant graviton configurations in backgrounds which are not linear
 perturbations to Minkowski, like the M-theory backgrounds
 $AdS_4\times S^7$ and $AdS_7\times S^4$ \cite{JL2,JLR3}.
In all cases perfect agreement with the
 description of \cite{GST,GMT} has been found in the limit of large number of gravitons, in
 which the commutative configurations of \cite{GST,GMT} become an increasingly better
 approximation to the non-commutative microscopical configurations \cite{myers}.
 
 In the next section we will use this action 
 to describe gravitational waves propagating in
 the maximally supersymmetric pp-wave background of M-theory. 
 In order to find the quadrupolar coupling to the 6-form potential of this
 background we will need to interchange the direction of propagation of the waves
 (the Killing direction in the action) with a compact direction that has to do with
 the $U(1)$ decomposition of the 5-sphere contained in the background as an $S^1$ bundle over
 the two dimensional complex projective space, $CP^2$.
 
 \section{pp-wave Matrix models from point-like gravitons}
 
 \subsection{The BMN Matrix model with coupling to the six-form potential}

The BMN Matrix model gives the dynamics of DLCQ M-theory 
 in its maximally supersymmetric pp-wave background along the direction $x^-\sim x^-+2\pi R$, in the
 sector with momentum $2p^+=-p_-=N/R$.
 In this section
 we are going to show that the same Matrix
 model, plus a coupling to the six-form potential, arises from the action (\ref{11daction})
 when the gravitons propagate in the pp-wave background 
 along the $x^-$ direction. More details about this construction can be found in \cite{LR}.
 
 To do this it is convenient to describe the 5-sphere contained in the background
 as an $S^1$ bundle over $CP^2$, and to introduce adapted coordinates to the
 $U(1)$ isometry associated to the $S^1$.  The $CP^2$ is most conveniently  defined for our purposes as the submanifold of $\R^8$ determined by the set of (four independent) constraints
 $\sum_{a=1}^{8}z_a z_a =1$, 
 $\sum_{b,c=1}^8 d^{abc}z_b z_c=\frac{1}{\sqrt{3}}z_a$
 where $\{z_1,\dots ,z_8\}$ parametrize a point in $\R^8$ (see for instance \cite{ABIY}). 
 
 Let us take $k^\mu=\delta^\mu_\chi$ in the action (\ref{11daction}), 
 with $\chi$ the coordinate adapted to the $U(1)$ fibre.
 Using this coordinate and the Cartesian coordinates $\{z_1,\dots ,z_8\}$,
 embedding the $CP^2$ in $\R^8$,
 the background metric and potentials read
 \begin{eqnarray}
 \label{backadap}
 &&ds^2=-4dx^+dx^- -\Bigl[(\frac{\mu}{3})^2(x_1^2+x_2^2+x_3^2)+
 (\frac{\mu}{6})^2 y^2\Bigr]
 (dx^+)^2 \nonumber\\
 &&\hspace{0.7cm}+dx_1^2+dx_2^2+dx_3^2+dy^2+y^2[(d\chi-A)^2+dz_1^2+\dots +dz_8^2]\, ,
 \nonumber\\
 &&C^{(3)}_{+ij}=\frac{\mu}{3}\epsilon_{ijk}x^k\, ,  i,j=1,2,3\, , \quad
 C^{(6)}_{+\chi abcd}=\frac{\mu}{3}y^6 f^{[abe} f^{cd]f} z^e z^f\, ,
 a,b,c,d=1,\dots 8\, ,
 \end{eqnarray}
 where $f^{abc}$ are the structure constants of $SU(3)$.  
 Note that the choice of adapted coordinates to the $U(1)$ isometry in the
 decomposition of the 5-sphere as an $S^1$ bundle over $CP^2$ reduces the 
 explicit invariance
 of the 5-sphere from $SO(6)$ to $SU(3)\times U(1)$ \footnote{The whole invariance under 
 $SO(6)$ should however still be present in a non-manifest way.}. Therefore the
 background is manifestly invariant under $U(1)^2\times SO(3)\times SU(3)\times U(1)$.
 Taking light-cone gauge, $x^+=t$, and the waves propagating along the $x^-$ direction we find (we have denoted the non-Abelian transverse scalars by Capital letters)
  \begin{eqnarray}
 \label{BIaction2}
 S&=&-\int dx^+ {\rm STr}
 \Bigl\{\frac{1}{y}\sqrt{\beta+4\dot{x}^- -\dot{X}^2-\dot{y}^2-
 y^2\dot{Z}^2}\,\Bigl(\unity-\frac{y^2}{4}[X,X]^2-\frac{y^6}{4}[Z,Z]^2\Bigr)
 \Bigr\}\nonumber\\
&&-\frac{\mu}{3}\int dx^+ {\rm STr} \Bigl\{ -i\epsilon_{ijk}X^k X^j X^i+
 \frac12 y^6 f_{[abe}f_{cd]f} Z^d Z^c Z^b Z^a Z^e Z^f\Bigr\}\, ,
\end{eqnarray}
where $\beta=[(\frac{\mu}{3})^2(X_1^2+X_2^2+X_3^2)+(\frac{\mu}{6})^2 y^2]$, 
$y$, the radius of the
 5-sphere, is taken to be commutative, consistently with the invariance of the background, and
 $\unity-\frac{y^2}{4}[X,X]^2-\frac{y^6}{4}[Z,Z]^2$ arises as the 
 expansion of the square
 root of the determinant of $Q$ up to fourth order in the embedding scalars\footnote{The
 same approximation is taken inside the square root in (\ref{BIaction2}). Note that
 this is the usual approximation taken in non-Abelian BI actions \cite{myers},
 which is valid when the non-Abelian action is good
 to describe the system of waves, that is, when the waves are distances away less than
 the Planck length (in our units $l_p=(\sqrt{2\pi})^{-1}$.}.

The most general non-commutative ansatz compatible with the symmetry of the background is to take 
$X^i=\frac{r}{\sqrt{C_N}}J^i$, $ i=1,2,3$, and
 $ Z^a=\frac{1}{\sqrt{C_N}}T^a$, $a=1,\dots ,8$,
 where the $J^i$ (the $T^a$) form an $N\times N$ representation of $SU(2)$ ($SU(3)$) and $C_N$ is the quadratic Casimir of the group in this
 representation. With this ansatz and Legendre transforming $\dot{x}^-$ to $p_-$ we 
 find\footnote{Note that the action (\ref{BIaction2}) describes gravitons which, by construction, carry $N$ units of momentum
 charge in the $\chi$ direction. 
 This momentum charge has to
 be set to zero in order to describe the sector of the theory with only
 light-cone momentum. How to do this consistently without putting to zero the number of gravitons, given also by $N$, is explained in \cite{LR}. We refer the reader to that reference for the details.}
  \begin{eqnarray}
 \label{newaction}
 H&=&-\int dx^+ {\rm STr}\Bigl\{ \frac{1}{4R}(\dot{X}^2+\dot{y}^2+y^2
 \dot{Z}^2)-\frac{1}{4R}(\frac{\mu^2}{9}X^2+\frac{\mu^2}{36}y^2)
 +\frac12 R [X,X]^2\nonumber\\
 &&-\frac{1}{16} R \,y^{10} [Z,Z]^4+i\frac{\mu}{3}
 \epsilon_{ijk}X^kX^jX^i-\frac{\mu}{6}y^6 f_{[abe}f_{cd]f} Z^d Z^cZ^b
 Z^aZ^e Z^f \Bigr\}\, .
 \end{eqnarray}
This Hamiltonian shares with the BMN Matrix model the $\R^3$ part of the geometry, whereas the $\R^6$ part  is described in a different system of coordinates, in which the $SO(6)$ invariance is reduced to $SU(3)\times U(1)$. What is most interesting is that we find a new coupling to the six-form potential of the background. This coupling supports a fuzzy 5-sphere giant graviton solution with the right radius in the large $N$ limit. 

The fuzzy 5-sphere solution is realized as an $S^1$ bundle over a fuzzy $CP^2$. The same kind of fuzzy spheres arise in the microscopical description of the 5-sphere giant graviton solutions of the 
 $AdS_4\times S^7$ and
$AdS_7\times S^4$ backgrounds \cite{JLR3}.
As in those cases the fuzzy $CP^2$ is constructed by making non-commutative the $z^a$ coordinates that parametrize $\R^8$. The set of four independent constraints is realized at the level of matrices by choosing the generators of 
 $SU(3)$ in the $(n,0)$ or $(0,n)$
irreducible representations.
 In these representations
$Z^a=\frac{1}{\sqrt{\frac13 n^2+n}}T^a$.

Substituting this ansatz in the Hamiltonian and taking
 $r=X^i=0$,  $i=1,2,3$, and $y$ and $Z^a$ time independent one gets
  \begin{equation}
 \label{HM5}
 H=\int dx^+ \frac{N}{R}y^2\Bigl( \frac{\mu}{12}-\frac{Ry^4}{4(n^2+3n)}
 \Bigr)^2\, .
 \end{equation}
Minimizing this expression we find two zero light-cone energy solutions.
One for $y=0$, 
the point-like graviton, and another one for
 $y=(\frac{\mu(n^2+3n)}{3R})^{1/4}$,
 which corresponds to the giant graviton solution. 
 Taking into account that the dimension of the $(n,0)$ and $(0,n)$
 representations is 
$N=\frac{(n+1)(n+2)}{2}$
 and that $p_-=-N/R$, we have that for large $N$,
$y\sim (-\frac23 \mu p_-)^{1/4}$
 which is the radius of the classical 5-sphere giant graviton solution of the pp-wave background (see \cite{LR}). Furthermore, in this limit the microscopical and macroscopical Hamiltonians agree exactly. This is an important check for the validity of the Matrix model given by
 (\ref{newaction}).

\subsection{The Type IIB pp-wave Matrix model}

A Type IIB pp-wave Matrix model can also be derived from the action for coincident gravitons constructed in \cite{JLR}. This action is very similar to the one describing M-theory gravitons that we have reviewed in the previous section. The main difference is that it contains a second isometric direction which allows to couple the 4-form RR-potential dielectrically. This action is particularized to the Type IIB background identifying $x^+$ with the worldline time, and taking the gravitons propagating in the $\psi$ direction\footnote{This is equivalent to taking the gravitons with momentum $p_-$, as explained in \cite{LR2}.}. The second isometric direction present in the action is identified with a $\Z_2$ symmetric combination of the two fibers associated to the two $S^1$'s  in the parametrization of the two transverse 3-spheres of the pp-wave background as $S^1$ bundles over $S^2$. Using then Cartesian coordinates to describe the 2-spheres one finds non-vanishing dielectric couplings of the RR 4-form potentials, which allow the existence of
zero energy solutions corresponding to expansions of the gravitons into the two 3-spheres
contained in the geometry. The details of this construction can be found in \cite{LR2}. Here we simply quote the final Matrix model derived in that reference:
 \begin{eqnarray}
 \label{IIBMatrix}
 H&=&-\int dx^+{\rm STr}\Bigl\{ \frac{1}{2R}\Bigl(\dot{r}^2+\dot{y}^2+
\frac{r^2}{4}\dot{X}^2+\frac{y^2}{4}\dot{Z}^2\Bigr)-
\frac{\mu^2}{2R}(r^2+y^2)+\nonumber\\
 &&+\frac{1}{256}R(r^2+y^2)\Bigl(r^4[X,X]^2+
y^4[Z,Z]^2+2r^2 y^2 [X,Z]^2\Bigr)+\nonumber\\
&&-i\frac{\mu}{8}r^4\epsilon_{ijk}
 X^iX^jX^k-i\frac{\mu}{8}y^4\epsilon_{abc}Z^aZ^bZ^c\Bigr\}
 \end{eqnarray}
This Type IIB Matrix theory is a $U(N)$ gauge theory built up with six
non-Abelian scalars, $X^i$, $i=1,2,3$, and $Z^a$, $a=1,2,3$ plus two
Abelian ones, $r$ and $y$, which are the radii of the two 3-spheres. The gauge field is set to zero through the gauge
fixing condition $A_\tau=0$.
In these coordinates the explicit symmetry
of the background is reduced to $(SO(3)\times U(1))^2$.

A non-trivial check of the correctness of this Matrix model 
is that it supports
fuzzy 3-sphere solutions which agree exactly,
in the limit of large number of gravitons, with the classical 3-spheres
of \cite{GST,SS}. Note that the 3-sphere giant graviton expanding in the
spherical part of the geometry \cite{GST} and the one expanding in the $AdS$ part 
\cite{GMT,HHI} of the $AdS_5\times S^5$ spacetime are mapped under Penrose limit into the same
type of solution, a fact that is reflected in the action through the
$\Z_2$ symmetry 
$r\leftrightarrow y$, $X\leftrightarrow Z$.
Let us consider for instance
the dual giant graviton solution, i.e. the one expanding into the (Penrose limit
of the) $AdS$ part of the geometry. The fuzzy 3-sphere ansatz is given by:
$r={\rm constant}$, $ y=Z^a=0$, $a=1,2,3$,
$X^i=\frac{1}{\sqrt{N^2-1}}J^i$, $ i=1,2,3$,
where $J^i$ are $SU(2)$ generators in an $N$ dimensional 
representation.
That is, we define the fuzzy
3-sphere as an $S^1$ bundle over a 
fuzzy 2-sphere.
Substituting this ansatz in (\ref{IIBMatrix}) we get
\begin{equation}
H=-p_-\, \frac{r^2}{2}\Bigl(\mu+\frac{2\pi^2 T_3 r^2}{p_-}\Bigr)^2\, .
\end{equation}
Therefore, the radius of the giant graviton solution is given by
$r^2=\frac{4\mu\sqrt{N^2-1}}{R}$, which agrees exactly with the macroscopical solution of \cite{GST,SS}. 

Our Matrix model is a one dimensional gauge theory which could be a candidate for the
holographic description of strings in the pp-wave background.
There is however a
second candidate for this holographic description, which is the Tiny Graviton Matrix theory of \cite{SJ}.   The obvious difference between both models is that
our Matrix
model does not depend on the Matrix ${\cal L}_5$, 
which lacks a direct physical interpretation. However this happens at
the expense of losing the explicit
$SO(4)\times SO(4)$ symmetry of the transverse space, and of the Matrix model in \cite{SJ}.
The existence of these two different Matrix models for the Type IIB pp-wave background
 could be related to the fact that there is no unique way to quantize diffeomorphisms in a 3-sphere. Therefore one could expect different gauge theories with the right continuum limit. We refer the reader to reference \cite{LR2} for a more extensive discussion on the differences between both Matrix models.

\subsection*{Acknowledgements}
Y.L. would like to thank the organizers of the 2nd workshop of the TMR network "Constituents, Fundamental Forces and Symmetries of the Universe''  for a very nice workshop and for giving her the opportunity to present this material.
The work of Y.L. has been
partially supported by CICYT grant BFM2003-00313 (Spain), and by the European
Commission FP6 program MRTN-CT-2004-005104, in which she is associated
to Universidad Aut\'onoma de Madrid. D. R.-G. is partially supported by the Fulbright-MEC fellowship grant FU-2006-07040 (Spain).


\begin{thebibliography}{[00]}

\bibitem{BMN}
D. Berenstein, J. Maldacena, H. Nastase, JHEP \textbf{0204},  013 (2002),
hep-th/0202021.

\bibitem{DSV}
K. Dasgupta, M.M. Sheikh-Jabbari, M. Van Raamsdonk, JHEP \textbf{0205}, 
056 (2002), hep-th/0205185.

\bibitem{SY}
K. Sugiyama, K. Yoshida, Nucl. Phys. \textbf{B644}, 128 (2002), hep-th/0208029.

\bibitem{DMS}
S.R. Das, J. Michelson, A.D. Shapere, Phys. Rev. \textbf{D70}, 026004 (2004),
hep-th/0306270.

\bibitem{SJ}
M.M. Sheikh-Jabbari, JHEP \textbf{09}, 017 (2004), hep-th/0406214.

\bibitem{Bonelli}
G. Bonelli, JHEP \textbf{08}, 022 (2002), hep-th/0205213.

\bibitem{SY2}
Y. Sekino, T. Yoneya, Nucl. Phys. \textbf{B619}, 22 (2001), hep-th/0108176.

\bibitem{Gopakumar}
R. Gopakumar, Phys. Rev. Lett. \textbf{89}, 171601 (2002), hep-th/0205174.

\bibitem{SS}
D. Sadri, M.M. Sheikh-Jabbari, Nucl. Phys. \textbf{B687}, 161 (2004),
hep-th/0312155.

\bibitem{JL2}
B. Janssen, Y. Lozano, Nucl. Phys. \textbf{B658}, 281 (2003), hep-th/0207199.

\bibitem{LR2}
Y. Lozano, D. Rodr\'{\i}guez-G\'omez, JHEP \textbf{0608}, 022 (2006),
hep-th/0606057.

\bibitem{JL1}
B. Janssen, Y. Lozano, Nucl. Phys. \textbf{B643}, 399 (2002), hep-th/0205254.

\bibitem{JLR3}
B. Janssen, Y. Lozano, D. Rodr\'{\i}guez-G\'omez, Nucl. Phys. \textbf{B712}, 
371 (2005), hep-th/0411181.

\bibitem{GST}
J. McGreevy, L. Susskind, N. Toumbas, JHEP \textbf{0006}, 008 (2000),
hep-th/0003075.

\bibitem{GMT}
M.\,T. Grisaru, R.\,C. Myers, O. Tafjord, JHEP \textbf{0008}, 040 (2000),
hep-th/0008015.

\bibitem{myers}
R.C. Myers, JHEP \textbf{9912}, 022 (1999), hep-th/9910053.

\bibitem{LR}
Y. Lozano, D. Rodr\'{\i}guez-G\'omez, JHEP \textbf{0508}, 044 (2005),
hep-th/0505073.

\bibitem{ABIY}
G. Alexanian, A.P. Balachandran, G. Immirzi, B.~ Ydri, J. Geom. Phys. \textbf{42},
28 (2002), hep-th/0103023.

\bibitem{JLR}
B. Janssen, Y. Lozano, D. Rodr\'{\i}guez-G\'omez, Nucl. Phys. \textbf{B669}, 363 (2003),
hep-th/0303183.

\bibitem{HHI}
A. Hashimoto, S. Hirano, N. Itzhaki, JHEP \textbf{0008}, 051 (2000),
hep-th/0008016.




\end{thebibliography}
\end{document}